\documentclass[prb,aps, superscriptaddress, reprint, amsmath,amssymb, showpacs]{revtex4-1}
\usepackage{graphicx}
\usepackage{dcolumn}
\usepackage{bm}
\usepackage[%
colorlinks=true,
urlcolor=blue,
linkcolor=blue,
citecolor=blue
]{hyperref}
\begin{document}

\title{Low compressible BP$_3$N$_6$}

\author{George S. Manyali}
 \email{gmanyali@mmust.ac.ke}
\affiliation{
Computational and Theoretical Physics Group, 190-50100, Kakamega, Kenya.
}
\affiliation{
	Department of Physics, Masinde Muliro University of Science and Technology,
	190-50100, Kakamega, Kenya} 
\affiliation{Department of Physical sciences,
	Kaimosi Friends University College,
	385-50309, Kaimosi, Kenya}
\author{James Sifuna}
\affiliation{
	Computational and Theoretical Physics Group, 190-50100, Kakamega, Kenya.
}
\affiliation{ Department of Natural Sciences, 
	The Catholic University of Eastern Africa,
	62157 - 00200, Nairobi, Kenya}
\affiliation{ Materials Modeling Group, 
	Department of Physics and Space Sciences,
	The Technical University of Kenya,
	52428-00200, Nairobi, Kenya}

\date{\today}

\renewcommand{\vec}[1]{\mathbf{#1}}

\begin{abstract}
Using first principles calculation, the structural and mechanical properties of BP$_3$N$_6$ which adopts an orthorhombic structure with space group Pna2$_1$ (no. 33), were determined at three different pressure values (0, 20 and 42.4~GPa). The nine independent elastic constants meet all necessary and sufficient conditions for mechanical stability criteria for an  orthorhombic crystal. BP$_3$N$_6$ show strong resistance to volume change hence a potential low compressible material. The Vicker's hardness of BP$_3$N$_6$ was found to range between 49-51~GPa for different external pressures imposed on the crystal. These high values of Vicker's hardness implies that BP$_3$N$_6$ is a potential superhard material.
\end{abstract}


\maketitle

\section{Introduction}
\label{intro}
Hardness is an important property that determines many of the technological applications of materials\cite{lyakhov2011evolutionary}. Designing materials based on first principles approach, synthesize and then characterization of these materials is of great interest to both theoretical and experimental material scientists. Well known superhard materials such as diamond and boron nitride display strong covalent bonds, low compressibility and high wear resistance. Applicability of these materials at high-pressure is limited and therefore there is need for new materials that would work perfectly in those harsh conditions. For many years now search for new superhard materials has been an order of the day. This search is unlikely to end any time soon due to the fact that diamond remains to be the hardest known material made of a single light element called carbon. Combination of carbon and other light elements such as nitrogen have shown promising results from the theoretical perspective\citep{tang2015first,manyali2012ab,manyali2013ab,manyali2013computational,manyali2015first} but most of the predicted materials have not been realized experimentally.  

The present study has been prompted by the recent explorative investigation of phosphorus nitrides by Vogel \textit{et.al.} \cite{vogel2018united,vogel2019boron} who synthesized a  high-pressure polymorph of boron phosphorus nitride ($\beta$-BP$_3$N$_6$). This phase adopts an orthorhombic structure with space group Pna2$_1$ (no. 33). The $\beta$-BP$_3$N$_6$ phase is characterized with octahedral coordinated phosphorous (P) atoms in a distorted hexagonal closed-packing of nitride anions. This phase was obtained at high-pressure experiment having transformed from $\alpha$-BP$_3$N$_6$ at a pressure of about 42~GPa. To our knowledge, no previous first-principles calculations have been carried out to investigate the mechanical stability and the hardness of this new phase of $\beta$-BP$_3$N$_6$ at different pressure. 

Mechanical stability is a concept based the elastic constants of a single crystal sample of a material. The elastic constants determine the response of the crystal to external forces that can be characterized in forms of bulk modulus, shear modulus, Young’s modulus, and Poisson’s ratio. All these properties can be obtained from density functional theory (DFT)\cite{hohenberg1964inhomogeneous,kohn1965self} which has become an essential tool for designing novel superhard materials .

In this paper, we first bench marked our study with previous theoretical studies on the lattice parameters, elastic constants, bulk and shear moduli of a hypothetical superhard carbon mononitride (Pnnm-CN)\cite{tang2015first}. This compound adopts an orthorhombic crystal structure of space group Pnnm (no. 58) and its Vicker's hardness has been predicted to be above 60 GPa. Once satisfied with Pnnm-CN results we went ahead to investigate the lattice parameters, elastic constants, bulk modulus,  shear modulus, Poisson's ratio, Young’s modulus, Vicker's hardness and mechanical stability of BP$_3$N$_6$ at different pressure. Both Pnnm-CN and BP$_3$N$_6$ adopts an orthorhombic structure with dense network of covalent bonds which are precursors for low compressibility.

The remaining parts of this paper are arranged as follows: in section \ref{sec:2}, we give a brief outline of the computational details. Results are shown and discussed in section \ref{sec:3}. Finally in section \ref{sec:4}, we provide a conclusion.

\section{Computational details}
\label{sec:2}

Calculations in the present work were performed using Density Functional Theory (DFT)\cite{hohenberg1964inhomogeneous,kohn1965self} within the Perdew–Burke Ernzerhof (PBE)\cite{perdew1996generalized} generalized gradient approximation and plane waves basis sets as implemented in the Quantum ESPRESSO package\cite{giannozzi2009quantum}. All pseudopotentials were obtained from the pslibrary1.0 maintained by Dal Corso\cite{dal2014pseudopotentials}. The cell volume and ions were optimized at three different pressure values i.e 0~GPa, 20~GPa and 42.4~GPa. The k-point sampling\cite{monkhorst1976special} was performed on 6x4x4 grid while the kinetic energy cut-off was set to 43 Ry. It is well known that PBE functional overestimates the lattice parameter and consequently affects not only the volume and the density of the compound but also the mechanical properties of the compound. An external pressure exerted on the crystal structure has also profound influence on the mechanical properties. Therefore, to get physically meaningful results from the calculations of elastic constants, we did not allow the ions to relax. This was necessary to maintain the strain on each ions at different pressure. The thermo\_pw package \cite{thermopw,urru2019spin} was used to run the elastic constants calculations. 

\section{Results and discussion}
\label{sec:3}
\subsection{Physical properties of Pnnm-CN at zero pressure}
\label{subsec:3:1}
In Table \ref{table:0cn} we have listed calculated physical properties of Pnnm-CN at zero pressure predicted using PBE functional. The calculated lattice constants produced from fitting total energy as a function
of volume to  Murnaghan\cite{murnaghan1944compressibility} equation of state  are compared to the corresponding theoretical results. In general, our calculated lattice constants and other properties shown in Table \ref{table:0cn} are in agreement with results of previous studies.  
\begin{table*}[tb]
	\centering
	\caption{Equilibrium structural parameters $a$ (\AA), $b$ (\AA), $c$ (\AA), elastic constants (GPa), bulk modulus (B), shear modulus G, $G/B$ ratio and Vickers hardness ($H_v$) in GPa  of Pnnm-CN calculated at zero pressure.}
	\label{table:0cn}
	\begin{tabular*}{\textwidth}{@{\extracolsep{\fill}}llllllllllllllllll@{}}
	\hline\noalign{\smallskip}
	&Pressure &a &b &c &$C_{11}$& $C_{22}$&$C_{33}$&$C_{44}$&$C_{55}$&$C_{66}$&$C_{12}$&$C_{13}$&$C_{23}$ &B&G&G/B&H$_v$\\ \noalign{\smallskip}\hline\noalign{\smallskip}
This work &0&5.340 &3.955&2.376 &499.2&635.8&1172.2&420.5&274.2&368.7&178.2&74.9 &138.9&328.7&322.6&0.98 &54.4\\
Others\cite{tang2015first}&     &5.335 &3.952 &2.374&506  &643  &1183  &442  &275  &372  &191 &80  &140 &336  &326 &0.97&62.5\\
	\noalign{\smallskip}\hline
	\end{tabular*}
\end{table*}
This results gave us confidence to study lattice parameters $a$ (\AA), $b$ (\AA), $c$ (\AA), volume $V$ (\AA$^3$)  and density $\rho$ (g$/$cm$^{3}$) of BP$_3$N$_6$ at zero and elevated pressures.
\subsection{Structural properties of BP$_3$N$_6$}
\label{subsec:3:2}
\begin{figure}[h!]
\centering
\resizebox{0.385\textwidth}{!}{%
\includegraphics{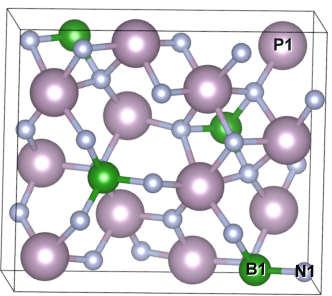}
}
\caption{The crystal structure of the primitive cell of BP$_3$N$_6$, where P are the large atoms labeled P1, B are medium-size atoms labeled B1,  whereas the smaller ones are N atoms labeled N1. The unit cell of contains 40 atoms and has a space group of Pna2$_1$ (no. 33).}
\label{fig:2}       
\end{figure}

Fig. \ref{fig:2} shows the crystal structure of the primitive cell of BP$_3$N$_6$ where P are the large atoms labeled as P1, B are medium-size atoms labeled as B1,  whereas the smaller ones are N atoms labeled as N1. The unit cell of contains 40 atoms and has a space group of Pna2$_1$ (no. 33).

Table.\ref{table:1} show the response of lattice parameter to different external pressure exterted on BP$_3$N$_6$. As the exerted pressure increase from zero to 20~GPa, the unit-cell contract by 1.87\%, 2.04\% and 1.79\% in $a,b$ and $c$ axes respectively. We see a lot more contraction in $b$ axis than in both $a$ and $c$ axes. This is a clear indication of existence of different bonding behavior along the axes. One would expect strong covalent bonding between boron and nitrogen atoms and weaker bonding character in phosphorus-nitrogen bond formation.
On increasing external pressure to 42.4~GPa, the unit-cell contract by 3.51\%, 3.78\% and 3.39\% in $a,b$ and $c$ axes respectively. Again, the $b$ axes display slightly more collapse in its distance than in both $a$ and $c$ axes.
At 42.4~GPa, the  calculated results are compared with corresponding experimental data from Vogel \textit{et. al.} \cite{vogel2019boron}. It is well known that PBE functonal overestimate the lattice constants and consequently the volume of the crystal structure. In this case, the predicted volume of BP$_3$N$_6$ is slightly higher that experimental values by 3.5\%. Furthermore, volumetric density of a material changes linearly with pressure and inversely with volume. At 42.4~GPa, the predicted density is lower than experimental value by 165g$mm^{-3}$. Generally, a good agreement is reached between computed and measured values.

\begin{table*}
\caption{Lattice parameters $a$ (\AA), $b$ (\AA), $c$ (\AA), volume $V$ (\AA$^3$)  and density $\rho$ (g$/$cm$^{3}$) of BP$_3$N$_6$  space group Pna2$_1$ (no. 33) calculated at different pressures in GPa.}
\label{table:1}
\begin{tabular*}{\textwidth}{@{\extracolsep{\fill}}lllllll@{}}
\hline\noalign{\smallskip}
	&pressure		& $a$   & $b$ &$c$ & $V$  & $\rho$ \\
		\noalign{\smallskip}\hline\noalign{\smallskip}
This work        &0   &4.212&7.789&9.063&297.3&4.193\\
                 &20  &4.133&7.630&8.900&280.7&4.442\\
                 &42.4&4.064&7.494&8.755&266.6&4.676\\
Expt.~\cite{vogel2019boron}&42.4&4.0115&7.411  &8.666  & 257.65& 4.841     \\		
		\noalign{\smallskip}\hline
	\end{tabular*}
\end{table*}


\subsection{Elastic stability of BP$_3$N$_6$}
\label{subsec:3:3}
Since 1954, the work of Born\cite{bornelastic} has been used to evaluate elastic stability of crystal structures. The  cubic crystals are easy to handle due their high symmetry. Efforts to apply the Born\cite{bornelastic} elastic stability criteria on lower symmetry crystals is reported in works of Ravindran \textit{et.al.} and that of Mouhat and co-author \cite{ravindran1998density,mouhat2014necessary}. These authors gave a precise description of the form it should take for lower symmetry crystal classes. They \cite{ravindran1998density,mouhat2014necessary} particularly considered orthorhombic crystal structure as one such system with lower symmetry. They described it as having nine independent elastic constants namely: $C_{11}$, $C_{12}$, $C_{13}$, $C_{22}$, $C_{23}$, $C_{33}$, $C_{44}$, $C_{55}$ and $C_{66}$. The mechanical stability of such orthorhombic crystal is achieved whenever the elastic constants satisfy the following necessary and sufficient conditions as prescribed by Born \cite{bornelastic}:
\begin{align}
	&C_{11} > 0 
	\nonumber\\
	&C_{11}C_{22} > C^2_{12} 
	\nonumber\\
	&C_{11}C_{22}C_{33}+2C_{12}C_{13}C_{23} - 
	C_{11}C^2_{23}-C_{22}C^2_{13}-C_{33}C^2_{12} > 0
	\nonumber\\
	&C_{44}>0
	\nonumber\\
	&C_{55}>0
	\nonumber\\
	&C_{66}>0
	\label{born1}
\end{align}

More work on orthorhombic crystal system has been illustrated by  Wen \textit{et.al.}\cite{wen2017ab}. These particular authors applied elastic stability conditions shown in Eq.~\eqref{born1} on orthorhombic TiAl alloy. They underscored the usefulness of Eq.~\eqref{born1} in checking stability of orthorhombic crystal system.
\begin{table*}
	\centering
	\caption{Elastic constants in GPa of BP$_3$N$_6$ calculated at different pressures.}
	\label{table:2}
	\begin{tabular*}{\textwidth}{@{\extracolsep{\fill}}lllllllllll@{}}
	\hline\noalign{\smallskip}
	& Pressure & $C_{11}$   & $C_{22}$   & $C_{33}$   & $C_{44}$&$C_{55}$&$C_{66}$&$C_{12}$&$C_{13}$&$C_{23}$ \\ \noalign{\smallskip}\hline\noalign{\smallskip}
BP$_3$N$_6$ &0    &818.3&720.1&788.6&296.9&302.5& 306.8&97.0&87.0 &141.4 \\
             &20  &972.5&876.2 &943.9 &345.1&360.6&367.4&164.2&151.5&198.6  \\
             &42.4&1103.0&1009.3&1073.1&385.2&409.5&419.1&224.1  &208.1 & 248.4 \\
	\noalign{\smallskip}\hline
	\end{tabular*}
\end{table*}
\\
In this work, the calculated values of the nine independent elastic constants at different values of pressure (0, 20 and 42.4~GPa) are listed in the Table \ref{table:2} and they satisfy the necessary and sufficient conditions prescribed for an orthorhombic system. All nine independent elastic constants increases monotonically with pressure. Already at zero pressure, BP$_3$N$_6$ depicts a material with large elastic constants which has a strong correlation with compressibility and hardness of a material. At elevated pressure, shorter covalent bonds will resist change in volume, a response that can be captured as enhanced elastic constants. At 42.4~GPa, BP$_3$N$_6$ has very large elastic constants comparable to that of diamond. We anticipate such large elastic constants will attract wide industrial application of BP$_3$N$_6$ and where possible a substitute to already existing hard materials.
It is noted that computing properties of a low symmetry system is tricky and therefore one has to be careful with the values of $C_{11}$, $C_{12}$, $C_{13}$, $C_{22}$, $C_{23}$, $C_{33}$, $C_{44}$, $C_{55}$ and $C_{66}$ as will always depend on the orientation of the unit cell. To our knowledge, there is no data on elastic constants to compare with our results and hence our work lays a foundation for future references.

From the single crystal elastic constants data\cite{wen2017ab}, the polycrystalline bulk modulus B and shear modulus G can be calculated by using the Voigt approximation \cite{voigt1928lehrbuch}; which describes the upper bound and the Reuss approximation \cite{reuss1929berechnung} that describes the lower bound. The Average of the two bounds is often described by the Hills approximation \cite{hill1952elastic}. Below are equations that relates the nine independent elastic constants to $B_V$ and $G_V$ in Voigt notation, whereas $B_R$ and $G_R$  represent Reuss notations.

\begin{align}
B_V = (C_{11} + 2C_{12} + 2C_{13} + C_{22} + 2C_{23} + C_{33})/9
\end{align}
\begin{align}
G_V = &(C_{11} - C_{12} - C_{13} + C_{22} - C_{23} + C_{33} + 3C_{44} 
\nonumber\\
&+ 3C_{55} + 3C_{66})/15 
\end{align}
\begin{align*}
\chi=&C_{13}(C_{12}C_{23} - C_{13}C_{22}) + C_{23}(C_{12}C_{13} - C_{23}C_{11})
\nonumber\\
& + C_{33}(C_{11}C_{22} - C_{12}^2)
\end{align*}
\begin{align}
B_R = &\chi(C_{11}(C_{22} + C_{33}- 2C_{23}) + C_{22}(C_{33} - 2C_{13})
\nonumber\\
 &-2C_{33}C_{12}+ C_{12}(2C_{23} - C_{12})+ C_{13}(2C_{12} - C_{13})
 \nonumber\\
 & + C_{23}(2C_{13} - C_{23}))^{-1}
\end{align}
\begin{align}
G_R = &15(4(C_{11}(C_{22} + C_{33} + C_{23}) + C_{22}(C_{33} + C_{13}) 
\nonumber\\
&+ C_{33}C_{12} - C_{12}(C_{23} + C_{12})-C_{13}(C_{12} + C_{13})
\nonumber\\
& - C_{23}(C_{13} + C_{23}))/\chi + 3(1/C_{44} + 1/C_{55} + 1/C_{66}))^{-1}
\end{align}
For simplicity, the averages of Voigt and Reuss approximations are presented as B and G for bulk and shear moduli respectively.

\begin{equation}
B=\frac{B_V+B_R}{2}   
\end{equation}

\begin{equation}          
G=\frac{G_V+G_R}{2}   
\end{equation}
The derived quantities from B and G are presented as Young's modulus (E) and and Poisson's ratio ($n$). 
\begin{equation}          
E =\frac{9BG}{3B + G}
\end{equation}
The Young's modulus is a measures for stiffness of a material while Poisson's ratio is often used to classify materials as either ductile or brittle\cite{haines2001synthesis}. This concept by Haines \textit{et. al.} will normally treat a brittle material as one having $n$ below 0.33 while a ductile material as one that has $n$ greater than 0.33. 
\begin{equation}       
n = \frac{3B - 2G}{6B + 2G} 
\end{equation}
On the other hand, Pugh's \cite{pugh1954xcii} definition of a ductile/brittle material is based on the $G/B$ ratio. Ductility is measured as a value of $G/B$ less than 0.5, while a material is rated brittle if $G/B$ goes beyond 0.5. Superhard phases of hypothetical $C_3N_4$ and diamond are good examples of brittle materials as demonstrated in the previous work of this author \cite{manyali2013ab}.   
\begin{table*}
	\centering
	\small
	\caption{bulk modulus ($B$), shear modulus ($G$), Young's modulus ($E$), and Poisson’s ratio ($n$ ) of BP$_3$N$_6$ in Hills', Voigt's and Reuss's approaches.}
	\label{table:3} 
	\begin{tabular*}{\textwidth}{@{\extracolsep{\fill}}lllllllllllllll@{}}
		\hline\noalign{\smallskip}
		&        & \multicolumn{3}{l}{Bulk modulus (GPa)} & \multicolumn{3}{l}{Young's modulus (GPa)} & \multicolumn{3}{l}{Shear modulus (GPa)} & \multicolumn{3}{l}{Poisson's ratio} \\ 
		\noalign{\smallskip}\hline\noalign{\smallskip}
		&Pressure&$B_V$&$B_R$&$B$&$E_V$&$E_R$ &$E$&$G_V$ & $G_R$ & $G$& $n_V$ & $n_R$ & $n$      \\ 
		\noalign{\smallskip}\hline\noalign{\smallskip}
 &0  &330.9&330.5&330.7&716.8&713.5&715.2&314.6&312.8&313.7&0.13897&0.14026&0.13961\\  
&20  &424.6&424.3&424.4&853.9&851.4&852.6&366.5&365.2&365.9&0.16485&0.16557&0.16521\\
&42.4&505.2&504.9&505.1&967.7&965.6&966.6&409.7&408.7&409.2&0.18077&0.18129&0.18103\\
		\noalign{\smallskip}\hline 
	\end{tabular*}
\end{table*}

  For this work, the calculated bulk modulus ($B$), shear modulus ($G$), Young’s modulus ($E$), and Poisson’s ratio ($n$ ) of BP$_3$N$_6$ in Voigt's and Reuss's approaches are listed in Table \ref{table:3}. The bulk modulus which is also referred to as incompressibility,  is a ratio that relates an applied constant stress to the
fractional volumetric change. BP$_3$N$_6$ has a bulk modulus of about 330~GPa comparable to many other low-compressible compounds such as SiO$_2$ with bulk modulus of 305~GPa, HfN with bulk modulus of 306~GPa, OsB$_2$ with bulk modulus of 297~GPa \cite{chung2008correlation,andrievskiy2007new,doughty1997hard,chen2011modeling}. The low compressibility of BP$_3$N$_6$ can be attributed to highly condensed network built up from the tetrahedra BN$_4$ and octahedra PN$_6$ coordination. Although it has been argued that incompressibility is a good indicator of hardness of a material, it is the shear modulus that is strongly associated with hardness. The shear modulus relates to how a material responds to change in shape at constant volume. Calculated shear modulus of BP$_3$N$_6$ is about 313~GPa and is in the same range as that of Pnnm-CN  reported as 322~GPa, see Tables \ref{table:3} and \ref{table:1} respectively. The current value of shear modulus (313~GPa) do not compete that of  diamond(546~GPa) as reported elsewhere\cite{manyali2012ab,manyali2013ab}.
It is worth noting that shear modulus increases with increasing pressure. This implies that a material has high resistance to shape change when it is under the influence of an external pressure.  
As indicated earlier, Young's modulus is a measures for stiffness of a material. It is derived from both shear and bulk moduli. Computed values of BP$_3$N$_6$'s Young's moduli for different pressure are presented in Tables \ref{table:3}. 
The calculated Poisson's ratio suggests that BP$_3$N$_6$ should be classified  as a brittle material  since the value of $n$ is less than 0.33 even at high pressure.  Similarly, the $G/B$ ratio calculated as 0.94 at zero pressure, 0.86 at 20 GPa and 0.81 at 42.4~GPa predicts BP$_3$N$_6$ as a brittle material with $G/B$ greater than 0.5. The values of $B/G$ for BP$_3$N$_6$ are shown in Table \ref{table:x2}. There is tendencies of BP$_3$N$_6$ to transform from brittleness to ductility when exposed high extremely high pressure as depicted by diminishing $G/B$ ratio at elevated pressure. This trend can also be interpreted to mean that BP$_3$N$_6$ will undergo pressure-induced phase transformation to lower symmetry crystal structures such as triclinic at extremely high pressures.  
\begin{table*}
	\centering
	\caption{The shear anisotropy factors ( $A_{\{100\}}, A_{\{010\}},  A_{\{001\}}$),  percent anisotropy factors of bulk (A$_B$)and shear (A$_G$) moduli, universal anisotropy factor $A^U$, $G/B$ ratio, Knoop hardness and Vickers hardness ($H_v$) in GPa  calculated for BP$_3$N$_6$ using PBE functionals.}
	\label{table:x2}
	\begin{tabular*}{\textwidth}{@{\extracolsep{\fill}}llllllllll@{}}
	\hline\noalign{\smallskip}
	& Pressure & A$_{\{100\}}$  & A$_{\{010\}}$& A$_{\{001\}}$& A$_B$ &A$_G$ &A$^U$ & $G/B$&$H_v$ \\ \noalign{\smallskip}\hline\noalign{\smallskip}
BP$_3$N$_6$ & 0&0.82&0.98& 0.91&0.05&0.287&0.02&0.94 &51.2    \\
            & 20&0.85&1.01&0.96&0.03&0.170&0.01&0.86 &50.1    \\
            & 42.4&0.87&1.03 &1.00 &0.02&0.120& 0.01&0.81 &49.7    \\

	\noalign{\smallskip}\hline
	\end{tabular*}
\end{table*}
Isotropy of crystal structure is very critical in characterization of any compound. The single crystal shear anisotropy factors A$_{\{100\}}$ in \{100\} planes, A$_{\{010\}}$ in \{010\} planes, and A$_{\{001\}}$ in \{001\} planes are defined as \cite{ravindran1998density}
\begin{align}
A_{\{100\}} = &4C_{44}/(C_{11} + C_{33} − 2C_{13}),\\
A_{\{010\}} = &4C_{55}/(C_{22} + C_{33} − 2C_{23}), \\
A_{\{001\}} = &4C_{66}/(C_{11} + C_{22} - 2C_{12}).
\end{align}
 The percentage anisotropy in compressibility  A$_B$ and shear A$_G$ is defined as \cite{chung1967elastic}
\begin{equation}
A_B =\frac{B_V - B_R}{B_V + B_R},\\
A_G =\frac{G_V - G_R}{G_V + G_R}.
\end{equation}
A zero value of A$_B$ and A$_G$ corresponds to elastic isotropy, while a value of 100\% corresponds to
the largest possible anisotropy. The universal anisotropy index (A$^U$) is defined as \cite{ranganathan2008universal}
\begin{equation}
A^U = \frac{B_V}{B_R}  + \frac{5G_V}{G_R} - 6.
\end{equation}
A nonzero value of A$^U$ is a measure of the anisotropy. The shear anisotropy factors ($A_{\{100\}}, A_{\{010\}}, A_{\{001\}}$), percent anisotropy factors of bulk (A$_B$) and shear (A$_G$) moduli, and universal anisotropy factor $A^U$ of BP$_3$N$_6$ are presented in Table \ref{table:x2}. For isotropic case, we expect A$_B$, A$_G$ and  $A^U$ to be zero and any deviation from zero is a clear indicator for anisotropy. Therefore, BP$_3$N$_6$ is anisotropic with highest deviations observed in both $A_{\{100\}}$ and A$_{\{001\}}$ directions. The $A_{\{100\}}$ direction is more isotropic than other directions.
In the introduction of this article, it was stated clearly that hardness is an important property that determines many of the technological applications of materials\cite{lyakhov2011evolutionary}. The Vicker's hardness in this work was estimated on the basis of the so called Chen model \cite{chen2011modeling} that has been widely used to predict  Vicker's hardness of a variety of crystalline  metals, insulators and semiconductors. The Chen's model is given as:
\begin{equation}
Hv =2(\frac{G^3}{B^2})^{0.585}-3.
\label{hvv}
\end{equation}
Where G and B are shear and bulk modulus respectively. The Vicker's hardness of BP$_3$N$_6$ is given in Table \ref{table:x2}. Clearly, BP$_3$N$_6$ has high Vicker's hardness close to that of Pnnm-CN and hence a potential superhard material. The strong resistance to change in volume and shape as a result of highly condensed orthorhombic structure explains this unexpected behavior in BP$_3$N$_6$. Right from large elastic constants to large bulk and shear moduli, to small values of  Poison's ratio ($\nu$) and large Vicker's harness, BP$_3$N$_6$ portray similar behaviour to well known superhard materials. However, pressure frustrates the hardness of BP$_3$N$_6$.

\section{Conclusion} 
\label{sec:4}
BP$_3$N$_6$ has been investigated at different pressure starting from zero pressure to a maximum of 42.4~GPa. Clearly,  the high value of bulk modulus indicate that BP$_3$N$_6$ is a low compressible material. This character emanates from the highly condensed orthorhombic structure built up from the tetrahedra BN$_4$ and octahedra PN$_6$ coordination. A  comprehensive analysis of elastic constants and its derived properties beside bulk modulus have been presented. BP$_3$N$_6$ displays high resistance to change in shape which greatly contributes to its high Vicker's hardness of about 51~GPa at zero pressure.  Overall, BP$_3$N$_6$ has similar behavior as those of well known superhard materials and hence the  conclusion that BP$_3$N$_6$ is a potential superhard material. These results provides guidance for further exploration of thermodynamic stability and many other properties of BP$_3$N$_6$ at different pressures. 
\section*{Acknowledgments}  
This work was financially supported by Kenya Education Network (KENET) through Computational Modeling and Materials Science (CMMS) Research mini-grants 2019. We acknowledge the Centre for High Performance Computing (CHPC), Cape Town, South Africa, for providing us with computing facilities.


\begin{thebibliography}{}
%
%

\bibitem{lyakhov2011evolutionary}Lyakhov, Andriy O., and Artem R. Oganov., Phys. Rev. B \textbf{84}, 9, (2011), 092103.
\bibitem{tang2015first}Tang, Xiao, Jian Hao, and Yinwei Li., Physical Chemistry Chemical Physics \textbf{17}, 41, (2015), 27821-27825.
\bibitem{manyali2012ab}Manyali, George S,Ab-initio study of elastic and structural properties of layered nitride materials, University of the Witwatersrand, Johannesburg,(2012).
\bibitem{manyali2013ab}Manyali George S and Warmbier, Robert and Quandt, Alexander and Lowther, John E, Comput. Mater. Sci., \textbf{69}, (2013) 299--303.
\bibitem{manyali2013computational}Manyali, George S and Warmbier, Robert and Quandt, Alexander, Comput. Mater. Sci., \textbf{79}, (2013) 710--714.
\bibitem{manyali2015first} G. S. Manyali, R. Warmbier and A. Quandt, Comput. Mater. Sci. \textbf{96}, (2015) 140-145.
\bibitem{vogel2018united}Vogel, Sebastian, Amalina T. Buda, and Wolfgang Schnick., Angewandte Chemie International Edition \textbf{57}, no. 40 (2018): 13202-13205.
\bibitem{vogel2019boron}Vogel, Sebastian \textit{et.al} Angewandte Chemie International Edition, (2019).
\bibitem{hohenberg1964inhomogeneous}P. Hohenberg and W. Kohn, Phys. Rev. \textbf{136},  (1964), B864.
\bibitem{kohn1965self}Kohn, Walter and Sham, Lu Jeu, Phys. Rev., \textbf{140}, 4A, (1965) A1133.
\bibitem{perdew1996generalized}J. P. Perdew, K. Burke, and M. Ernzerhof, Phys. Rev. Lett. 77, 3865 (1996).
\bibitem{giannozzi2009quantum}Giannozzi, Paolo \textit{et.al}, Journal of physics: Condensed matter, \textbf{21}, 39, (2009) 395502.
\bibitem{dal2014pseudopotentials}Dal Corso, Andrea, Comput. Mater. Sci., \textbf{95}, (2014), 337--350.
\bibitem{monkhorst1976special}H. J. Monkhorst and J. D. Pack, Phys. Rev. B 13, 5188 (1976).
\bibitem{thermopw}Corso, Andrea, Journal of Physics: Condensed Matter, \textbf{28}, 7, (2016), 075401.
\bibitem{urru2019spin} Palumbo, Mauro and Dal Corso, Andrea, physica status solidi (b), \textbf{254}, 9, (2017)
\bibitem{murnaghan1944compressibility}F. D. Murnaghan, Proc. Natl. Acad. Sci. USA 30, 244 (1944).
\bibitem{bornelastic}Born, Max and Kun Huang, Dynamical theory of crystal, (Clarendon press, 1954) 
\bibitem{mouhat2014necessary}Mouhat, Felix, and François-Xavier Coudert. Phys. Rev. B \textbf{90}, 22 (2014), 224104.
\bibitem{ravindran1998density} Ravindran, P and Fast, Lars and Korzhavyi, P A\_ and Johansson, B and Wills, J and Eriksson, O, Journal of Applied Physics, \textbf{84}, (1998) 4891--4904.
\bibitem{wen2017ab}Wen, Yufeng, Long Wang, Huilong Liu, and Lin Song, Crystals \textbf{7}, 2 (2017) 39.
\bibitem{voigt1928lehrbuch}Voigt, Woldemar. Lehrbuch der kristallphysik. \textbf{962} Leipzig: Teubner, 1928.
\bibitem{reuss1929berechnung}Reuss, A., Angew, Z. Berechnung del fliessgrenze von mischkristallen auf grund der plastizitatbedingung for einkristalle. Math. Mech. \textbf{9},(1929), 49–58.
\bibitem{hill1952elastic}Hill, R. The elastic behaviour of a crystalline aggregate. Proc. Phys. Soc. A, \textbf{65} (1953), 349–354.
\bibitem{haines2001synthesis} J. Haines, J.M. Leger, G. Bocquillon, Annual Review of Materials Research \textbf{31}, 1, (2001) 1–23
\bibitem{pugh1954xcii}S.F. Pugh. 1954  {\it Philos. Mag}. \textbf{45} 823–843.
\bibitem{chung2008correlation}H.Y. Chung, M.B. Weinberger, J.M. Yang, S.H. Tolbert, R.B. Kaner, Appl. Phys.
Lett. \textbf{92}, (2008) 261904.
\bibitem{andrievskiy2007new}R.A. Andrievskiy, in: Proceeding of the Seventeenth International Offshore and
Polar Engineering Conference, (2007).
\bibitem{doughty1997hard} C. Doughty, S.M. Gorbatkin, T.Y. Tsui, G.M. Pharr, D.L. Medlin, J. Vac. Sci.
Technol. A \textbf{15} (1997) 2623.
\bibitem{chen2011modeling} Chen, Xing-Qiu and Niu, Haiyang and Li, Dianzhong and Li, Yiyi, Intermetallics,
  \textbf{19}, (2011) 1275--1281.
\bibitem{chung1967elastic} Chung, DH and Buessem, WR, Journal of Applied Physics, \textbf{38}, (1967), 2010--2012. 
\bibitem{ranganathan2008universal}Ranganathan, Shivakumar I and Ostoja-Starzewski, Martin, Phys. Rev. Lett., \textbf{101}, (2008) 055504.
%

\end{thebibliography}
\end{document}